\documentclass[aps,prd,superscriptaddress,eqsecnum,amsfonts,showpacs,epsfig]{revtex4}
\usepackage{epsfig}

\newcommand{\be}{\begin{equation}}
\newcommand{\ee}{\end{equation}}
\newcommand{\bea}{\begin{eqnarray}}
\newcommand{\eea}{\end{eqnarray}}

\newcommand{\nn}{\nonumber}
\newcommand{\ep}{i\epsilon}

\begin{document}

%\preprint{ \parbox{1.5in}{\leftline{hep-th/0111433}}}

\title{Quark spectral functions from spectra of mesons and vice versa}

\author{V. \v{S}auli}

\email{sauli@ujf.cas.cz}
\affiliation{Department of Theoretical Physics, Institute of Nuclear Physics Rez near Prague, CAS, Czech Republic  }

\begin{abstract}
We extract the spectral function of quarks using the QCD functional formalism.The reliability of  approximations is controlled by the masses and decays of pseudoscalar mesons. For light quarks, we select the pion; for charm quarks, we choose the ground and excited  states of the $\eta_c$ meson. To this end, we solved the ladder-rainbow approximation of the spectral Dyson-Schwinger equations for quarks coupled to the Bethe-Salpeter equation for the pion and pseudoscalar charmonia. The obtained dynamical charm mass function varies from 1 to 1.5 GeV in the timelike domain, which is significant for narrow charmonium states. This yields a notable response in the bound state spectrum. We obtained purely continuous spectral functions and discussed their connection with confinement.
 \end{abstract} 
\maketitle

\section{Introduction}

  Knowledge of QCD correlation functions in the time-like momentum region is crucial for the first-principles determination of hadronic resonances and the understanding of hadron production \cite{s3,s4}.  In this regard, the analytical continuation of lattice data for propagators to timelike Minkowski space has been  challenged 
 \cite{DORS2020,LD2022}.   A complementary method to lattice reconstruction is the spectral functional formalism. In this approach, the analytical continuation is performed from the outset using a suitably truncated set of Dyson-Schwinger equations (DSEs), and the equations  are solved for spectral functions in Minkowski space. 
  Since the initial spectral study \cite{CO1982},  the  method has recently been recognized as valuable \cite{s1,s2,SCLK2019,s3,MS2021,HPW2022b} and  includes the topic of spectral renormalization - primary or secondary subtractions technique performed  at the timelike momentum scale. The  Yang-Mills sector of $SU(3)$ gauge theory was considered in Landau gauge \cite{HPPW2021,HPW2022} bringing new insights for a moderate coupling models. Furthermore, the first primitive solution \cite{s2}, shows the importance of transverse vertices in strongly coupled pure gluodynamics.    
 
This paper focuses on the quark propagator — particularly the heavy quark propagator — and its spectral function.  
It offers an alternative explanation for the heavy quarkonium mass spectra by accounting for the often overlooked running quark mass effect. Recall the behavior of the mass function at the naive threshold: the QCD quark mass grows with the timelike scale, which is important for quarkonium physics.  Thus, the heavier excited states are effectively composed of quarks that are heavier than those constituing the ground state.   
In other words, when retardation effects are correctly accounted for in the fully covariant approach of BSEs, the radiative corrections that are responsible for running masses provide the same quarkonium spectrum as is predicted by nonrelativistic quantum mechanics, which includes a phenomenological confining potential in addition to the expected Coulomb term. The nonrelativistic Schrödinger equation with a confining interaction was introduced historically in \cite{KOSU1974} and  used broadly  in  recent extended studies \cite{ZHCZ2025,GUGEBH2021,JUHUCHA2021,strange2021,BG2020,SG2020,BGPSS2019,SJS2018,LLMV2018,DLGZ2017,LMZV2015,BS2013,LZ2011}  
 and can effectively replace the quantum field theory effect of running mass.

To determine whether this scenario applies to heavy quarkonia, we employ the DSE for heavy quarks, coupled with the BSE for heavy quarkonia, as was done for quarkonia  \cite{HPGK2015} and many times for the lighter mesons \cite{MARO1997,Sa2008,CK2010,QCLRW2012,FSV2014,TGK2015,HGK2017,HGKL2017,YCKRSX2019} where the running quark mass is a criticaly needed phenomenon. However, we avoid introducing a covariantized form of the confining potential, as was done earlier by the authors in  \cite{saulieta,saulipsy}.  This would interfere badly with the similar effect stemming from the considered running quark mass. Interestingly, the pion and heavy quarkonia (pseudoscalar charmonia) can be described by nearly the same functional form of the DSE kernel.

In the next section, we describe the calculation scheme for the quark propagator and the BSE. 
Section III describes the method for BSE solution and for the extraction of the light quark spectral functions. 
Section IV presents the same but for the  pseudoscalar charmonium.
We conclude in the last section V.
 
 \section{Truncation of SDEs system}
 
For the sake of completeness, we will write down all the necessary equations here. 
The quark propagator S can be parametrized as follows:
\be   \
S(q,\mu)=[A(q^2,\mu)\not q- B_q(q^2,\mu)]^{-1}
\ee
where  $A,B$ are two scalar functions that completely characterize the quark propagator when the electroweak interaction is turned off.
The DSE for the quark propagator in the MOM renormalization scheme can be written as
 \bea  
A(q^2,\mu)&=&a(\mu)-Tr \frac{\not q}{4q^2}\Sigma_r(q)+ Tr \frac{\not q}{q^2}\Sigma_r(q)|_{q^2=\mu^2}
\nn \\
B(q^2,\mu)&=&b(\mu)+\frac{Tr}{4}\Sigma_r(q)- \frac{Tr}{4} \Sigma_r(q)|_{q^2=\mu^2}
\nn \\
 \label{gap}
\Sigma_r(q)&=&i\frac{4}{3}\int\frac{d^D k}{(2\pi)^D
} \gamma_{\mu} S(q-k) \Gamma_{\nu}(k,q) G^{\mu\nu}(k) \, ,
\eea
where $Tr$ is for the trace over the Dirac indices.
The inverse of $A$ is traditionally called the quark renormalization function. The 
rate $B/A$ represents the renormalization-scheme invariant  dynamical quark function $M(q^2)$. The MOM renormalization scheme is known to be particularly useful for  evaluating  spectral functions and provides a straightforward interpretation of the constants $a(\mu)$ and $b(\mu)$. These constants represent the values of the renormalized functions $A(\mu^2,\mu)$ and $B(\mu^2,\mu)$, respectively.  As in the \cite{s1} quark selfenergy $\Sigma_r$ has been regularized dimensionally 
before the intended subtractions.

The usual routine is to use the "large" spacelike renormalization scale when solving the DSE for the quark propagator to obtain the current quark mass.
However, this is an unfavorable choice when solving the DSE for the quark spectral function because it causes numerical convergence to fail.  
Hence, to renormalize the light quark propagator,  we  take $\Re a(\mu)=1$ and  $ \Re b(\mu)=300 MeV$ for a small time-like renormalization scale  $\mu^2=0.5 GeV^2$.
The imaginary parts of  the functions $A$ and $ B$ are  read from their upper cut values at $p^2+\ep$ in the complex plane. In other words, they are  not  arbitrary,  and their values must  be found as a solution. If one requires that all renormalization constants be real,  then only the real parts of the projected self-energy can be subtracted in Eq. (\ref{gap}) to ensure  that all renormalization constants are real-valued.  To find the absorptive parts, we use  the method as developed in the  paper \cite{s1}, i.e.,  we search for their values by scanning intervals and considering the optimal values as results.

In the Eq. (\ref{gap}) $G $ stands for the dressed gluon propagator and   $\Gamma(k,q)$ represents  the dressed quark-gluon vertex. Their color and Dirac indices are suppressed, and Lorentz indices are omitted unless necessary. Assuming that the full vertices have the same color structure as the classical ones yields the prefactor of $4/3$, which is explicitly shown.
The popular Ladder-Rainbow Approximation (LRA) is used and  for the product of the quark-gluon vertex $\Gamma$ and the  gluon propagator $G$  we take 

\be \label{kernel}
\Gamma_{\nu} G^{\mu\nu}(p)=\gamma_{\nu}N(\xi)\left[-g^{\mu\nu}+\frac{p^{\mu}p^{\nu}}{p^2}\right]\int do \frac{\rho_T(o)}{p^2-o+\ep}
 -\not p \frac{\xi g^2 p^{\mu} }{(p^2)^2} \, ,
\ee
where $\rho_T$ is the LRA spectral function- a slightly general version of the Landau gauge gluon spectral spectral function.  $N(\xi)$ stands for  momentum 
independent constant factor and $\xi$ stands for  the usual  gauge fixing parameter.

In the Landau gauge, the function $\rho_T$ would provide the effective QCD charge, which falls as $ln^{-1}[Q/\Lambda_{QCD}]$ . Although results already exist for heavy quarks in this approximation, solving the associated coupled DSE for the spectral function and BSE for mesons is extremely difficult.     
In this report, we do not solve the gluonic spectral DSE or use a running charge in order to reduce the number of numerical integrations. Instead, we have used a simplified foolowing fit for the spectral function
\be \label{prase}
\rho_T(o)=-\delta(o-m_g^2)+\delta(o-\Lambda^2) \, ,
\ee
with $m_g=0.6 GeV$ and $\Lambda=2 GeV$ as found in \cite{s2}. Let us mention here, that the undergoing Landau gauge study, which  uses more realistic running coupling, lifts the gluonic scales $m_g$ and $\Lambda$ significantly $m_g\simeq 1.5 GeV$, being thus very close to original estimate \cite{CO1982} for the  gluonic mass scale. 

 The parameters $N(\xi)$ as well as the gauge parameter $\xi$ were varied to find the solution which complies with
\

{\bf 1.} spectral property of the quark propagator
\

{\bf 2.}  meson properties.  
 \  

To ensure the first point, the   DSE is solved  for the  spectral functions $\sigma$ for which purpose 
we  follow  the method  established and  described in details in the paper \cite{s1}. 
The method provides a solution to the equation(\ref{gap}) in the entire Minkowski space, 
thereby confirming that the spectral representation for the quark propagator exists in the standard form:
\be  \label{spec1}
S(p,\mu)=\int_0^{\infty} d o \frac{\not p \sigma_v(o)+\sigma_s(o)}{p^2-o+\ep}\, \, .
\ee

Two introduced spectral function stands for the dirac and the scalar part of the quark propagator
\bea
S_v(p^2)&=&\frac{A(p^2)}{p^2 A^2(p^2)-B^2(p^2)}=\int_0^{\infty} d o \frac{ \sigma_v(o)}{p^2-o+\ep}
\\
 S_s(p^2)&=&\frac{B(p^2)}{p^2 A^2(p^2)-B^2(p^2)}=\int_0^{\infty} d o \frac{ \sigma_v(o)}{p^2-o+\ep} \, ,
\label{spec2}
\eea
 where we have introduced   another standard labeling  for two scalar propagator functions $S_v$ and $S_s$; 
obviously $S={\not{p}} S_v+S_s$  . 
Let us mention that spectral function, likewise the propagator, is a scheme dependent object.  In gauge theory like QCD, it nontrivialy 
 depends on the gauge fixing  parameter as well as on the renormalization point. We will not indicate these dependencies in a further text for brevity.

The desired analytical properties are  not automatically guaranteed in QCD  as suggested by lattice study \cite{BIT2020}  and the spectral representation becomes more complicated. 
We consider the spectral representation to be an approximation, and 
 we minimize the spectral deviation  introduced in \cite{s1,s2} in order 
to comply with the point 1. To satisfy point 2 simultaneously, the Bethe-Salpeter equation (BSE) for the pion must produce the correct values of the pion mass, $m_{\pi} = 140 MeV$, and the pionic decay constant, $f_{\pi} = 90 MeV$. This defines the parameter space for modeling the interaction kernel of our LRA, but it also limits the physical renormalization conditions for the quark propagators.  Notably, within the truncation presented here it was shown in \cite{s1}, that the celebrated dispersion relation form \cite{CAGA1961} for Vacuum Hadron Polarization can be obtained as well as  the resulting  dispersion relation for  the electromagnetic meson form factor \cite{s3} appears naturaly.

 \begin{figure}
\centerline{\includegraphics[width=8.6cm]{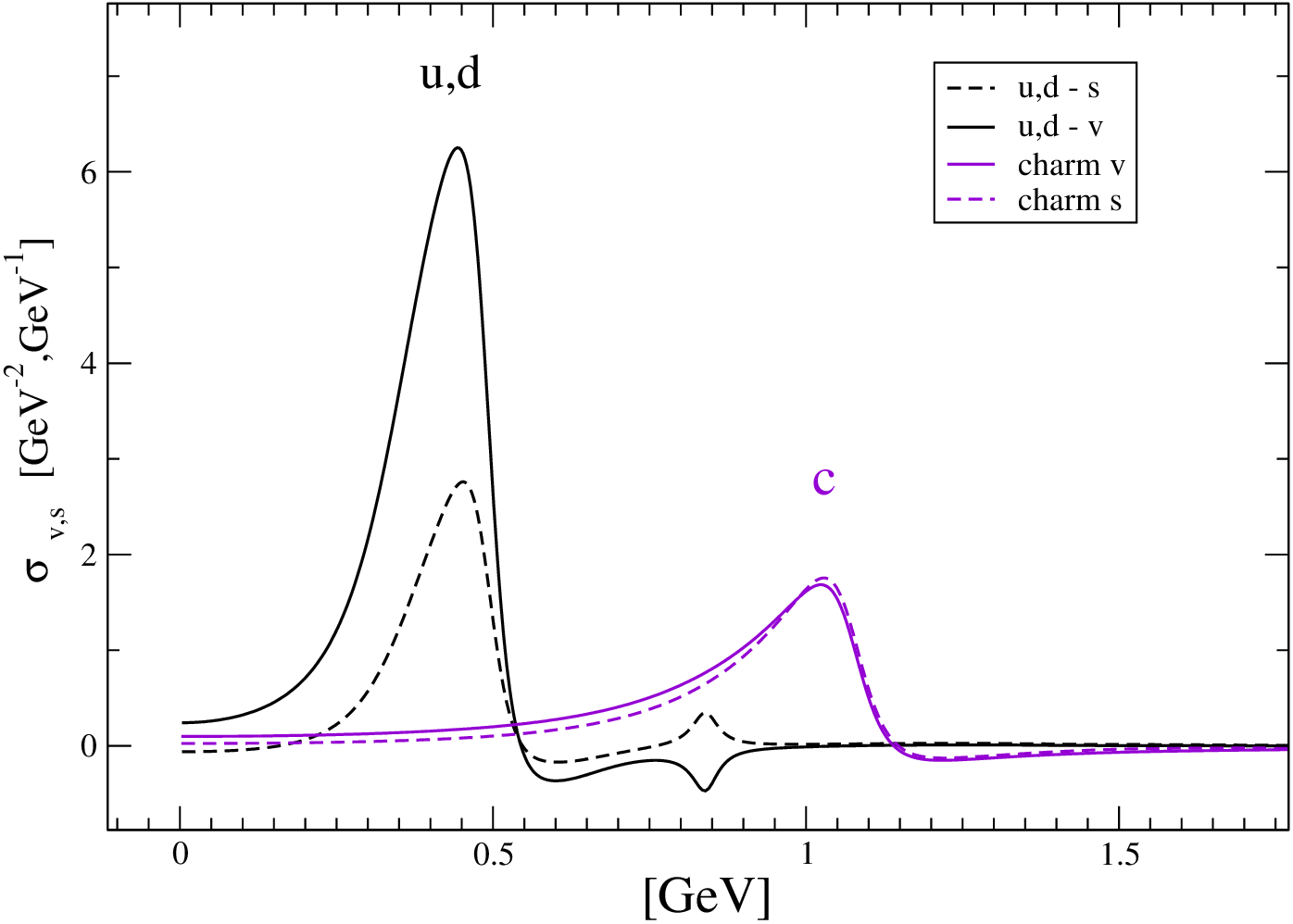}}
\caption{Quark spectral functions, solid line stands for the functions $\sigma_d$,  dashed line for $\sigma_s$ as plotted against  energy . 
The two blobs on the left are for light quarks and the blob on the right is for the charm quark.}
\label{jedna}
{\mbox{-------------------------------------------------------------------------------------}}
\end{figure}

\section{Results for the light quark spectral function from the pion properties}

The homogeneous BSE for mesons is a field-theoretical equation for the quark-antiquark bound state. In a dense notation it reads
\be
\Gamma=\int_k S\Gamma S K
\ee
where $S$ is the propagator of the bound state component, $\Gamma$
is the Bethe-Salpeter vertex function and $K$ is the quark-antiquark-quark-antiquark irreducible interaction kernel.
For the pion case, this kernel is taken within the same  approximation as for  the DSE in order to consistently preserve the  Goldstone character of the pseudoscalar mesons in the chiral limit.

To determine the spectrum, the single-component approximation works well 
 for the ground state \cite{FNW2008} as well as for the excited states, confirmed also by the charmonium study \cite{saulipsy}.
We used the dominant vertex component $\Gamma =\gamma_5 A$ and solved the BSE was solved using the eigenvalue method in complex momentum space.

To get the numerical solution, 
we introduce a suited auxiliary eigenvalue function, $\lambda(P, A)$. Then, after 3D angular integration and some trivial algebra, the BSE to be solved by method of iterations reads:
\bea \label{num}
A(p_E,P)=\lambda(P,A)\int\limits_{-\infty}^{\infty} dk_4 \int\limits_0^{\infty} d {\bf k} \frac{\bf k^2}{\bf p^2}  A(k_E,P)(S_v(k_+)S_v(k_-)(k_E^2+m_{\pi}^2)+S_s(k_+)S_s(k_-))[K_g(k,p)+K_{\xi}(k,p)] \, .
\eea

As  one can see, the rhs. of the BSE involves
 the product of the scalar functions
$S_v(k_+)S_v(k_-)$ and $S_s(k_+)S_s(k_-)$ evaluated at complex- valued momentum $k_{\pm}=k\pm P/2$. Here $k$ (or $k_E$) is a  relative real Euclidean momentum, while the total momentum  $P_E=(im_{\pi},0)$ in the rest frame of the pion. The propagator $S$ is  becoming complex for complex arguments. For an explicit evaluation,  we have used the spectral representation (\ref{spec2}) and  implemented an additional integration over the spectral variable in the BSE kernel.  However, note the integration factorizes and the   both  propagator  products  appearing in the Eq. (\ref{num}) stay real since both propagators  in the product are identical. The kernel $K_g$ and $K_{\xi}$ are usual logs stemming from the "gluon propagator" integrated over the 3d spacelike angles. Thus for instance 
\be 
K_{\xi}=C_A\frac{g^2\xi}{(16\pi^3)} \ln\frac{f+2{\bf k p}}{f-2{\bf k p}}
\nn
\ee 
with $f$ defined as $f=k_E^2+p_E^2+2 k_4p_4$.

\begin{figure}
\centerline{\includegraphics[width=8.0cm]{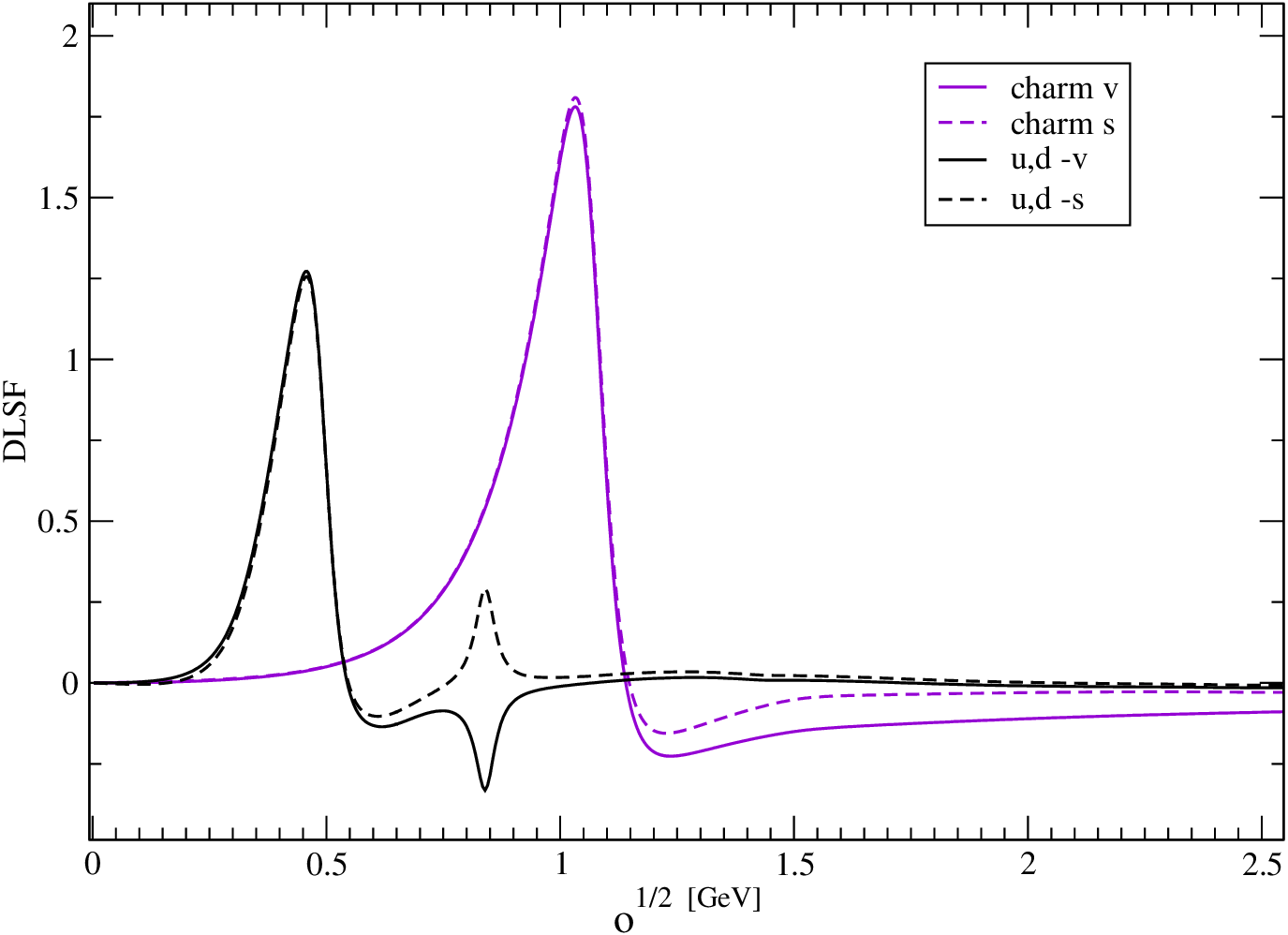}}
\caption{Dimensionless quark spectral functions $ o\sigma_v(o)$ (solid line)  and $\sqrt(o)\sigma_s(o)$ (dashed line)
 for the light and charm quark. At larger (smaller) energy scale the broad peak for the charm flavor  (u,d)  quark spectral function develops.}
\label{dva}
{\mbox{---------------------------------------------}}
\end{figure}

To solve the quark DSE, we assume that anomalous branch points emerge according to the Schwinger picture of confinement.
We found that the branch point is located at the origin of the complex plane  rather than at its perturbative value, which is equal to the sum of the pole masses, 
$m_g + m_p$. However, the pole mass, $m_p$, does not exist at all (in fact neither does  $m_g$ ).
The quark propagator has a branch cut along the entire positive real semiaxis of the $p^2$ variable.  
The resulting spectral functions for the light quarks are shown in the Figure \ref{dva}. The spectral function for the u quark is equal to that for the d quark because we work in the isospin limit and ignore the electromagnetic interaction.

The broad shapes of both functions $\sigma_{v,s}$ correspond to  confined objects- the light quark excitation is always confined inside the pion.
It is important to note that the  disappearance of the Dirac delta function is a  rigid property of the observed solution.
The singular peak reappears only when  the DSE kernel is made unphysically weak by hand. If one does this explicitly in the DSE for the quark propagator, the same should be done in the BSE. However, this immediately causes one to lose the pion properties of the pionic BSE. This reflects how confinement and chiral symmetry breaking are manifest in the spectral properties of the quark propagator.
Consequently, when the pole is absent from the quark propagators, color production thresholds are not generated.  
This behavior is intuitive and is provided by the solution.  There is no need to mimic it by hand, as was done in \cite{BFGIKL2010,GGIL2021,DDIL2022}
for the purpose of form factor calculations. We obtain confinement of quarks through the solution, both qualitatively and quantitatively.

The following rate of couplings and gauges 
\be \label{rate}
\frac{g^2\xi}{N(\xi)}=3 \, ; \frac{4N(\xi)}{3(4\pi)^2}=\frac{16}{3}
\ee
provides one particularly convergent solution in our LRA.
Recall that, as observed in \cite{sauliFBS}, extending to other spin mesons is not straightforward in the class of nontrivial covariant gauges. 
However here for pseudoscalar meson considered here,  the  deviation from assumed analyticity   as  established in \cite{s1,s2} seems to be vanishing. 
The  value $\sigma^2=10^{-6}$ has been achieved and appears to be limited only by the precision of the principal value integral that appears in its definition. 
Our truncation of  BSE/DSE  system do not provide excited states as solution of homogeneous equation,which is  accordance with  broad resonant character of experimentally known  $\pi$*.  Oppositely, as we describe in the following section, the formalism provides all narrow excited states of pseudoscalar charmonia.

The numerical codes used to solve the BSE are based on an iterative search. We scan the interval of expected mass solutions using 90 iterations for each meson mass.
After obtaining the solutions, we store the information and use it to find solutions with greater precision until the desired level of satisfaction is achieved.    
The final BSE and DSE search codes are available to the public \cite{mujweb}. 

\section{ c-quark spectral function from $\eta_c(n)$ quarkonia }

In this section we describe necessary but minimal  changes to get the numerical solution for the charmonium pseudoscalar system. 
The numerical solution is described below as well.

To find the $\eta_c$ and its excitations we  used the same functional form of  DSE/BSE interacting
kernel as for the previous calculation of the pion. Furthermore,  the dimensionless quantities  $g$ and $ \xi$ remain exactly the same as in the pion case.
According to the flavor non-universality of the quark-gluon vertex, which is a part of our effective coupling  a  certain changes of the interaction  is required  to comply with the   charm  mass scale. This softening is  achieved changing the renormalization scale as well as  by readjusting  the masses that characterize the 
kernel, which now takes the following values

 \be
m_g^c=0.433\,  GeV \, \, ; \, \, \Lambda^c =1.442 \,  GeV ;
\ee 
where we add subscript $c$ since a  charm flavour. 

We fixed the  renormalization function and the mass  such that $\Re a_c(\mu)=1$  and $\Re M_c(\mu)=\Re b_c(\mu)=0.94  GeV$
for very low, yet still time-like renormalization point  $\mu^2=0.13 GeV^2$. We also mimicked the effect that could come from higher order skeleton diagrams, such as cross boxes, which carry  total momentum dependence.
To do so, we insert the following prefactor to mimic the effect beyond LRA:
\be \label{pref}
f_{\eta}=\frac{1}{\sqrt{2}}\sqrt{1+\frac{M(\eta_c(2))^2}{P^2}}.
\ee
in rhs. of the BSE. 
Notably, the factor $f$  does not contribute to the ground state.

The coupled system of the DSE for the quark propagator and the BSE for pseudoscalar charmonia was solved in similar manner as for the pion.
However, the excited narrow states  were also obtained this time.
Quite interestingly, the expected "textbook" value of the charm mass
$M_c=1.5 GeV$ is reached approximately at  the maximum of the dynamical quark mass fucntion $M(p)=B(p)/A(p)$. 
The mass function in the  spacelike domain is smaller 
than $1GeV$.  Additionally, at the scale corresponding to the ground state,  the charm function mass function is smaller than expected from nonrelativistic studies. It approaches the value  $M(M_{\eta_c}/4)\simeq 1.1 GeV$ at the ground state mass scale.

\begin{figure}[htb]
\includegraphics[width=7.cm]{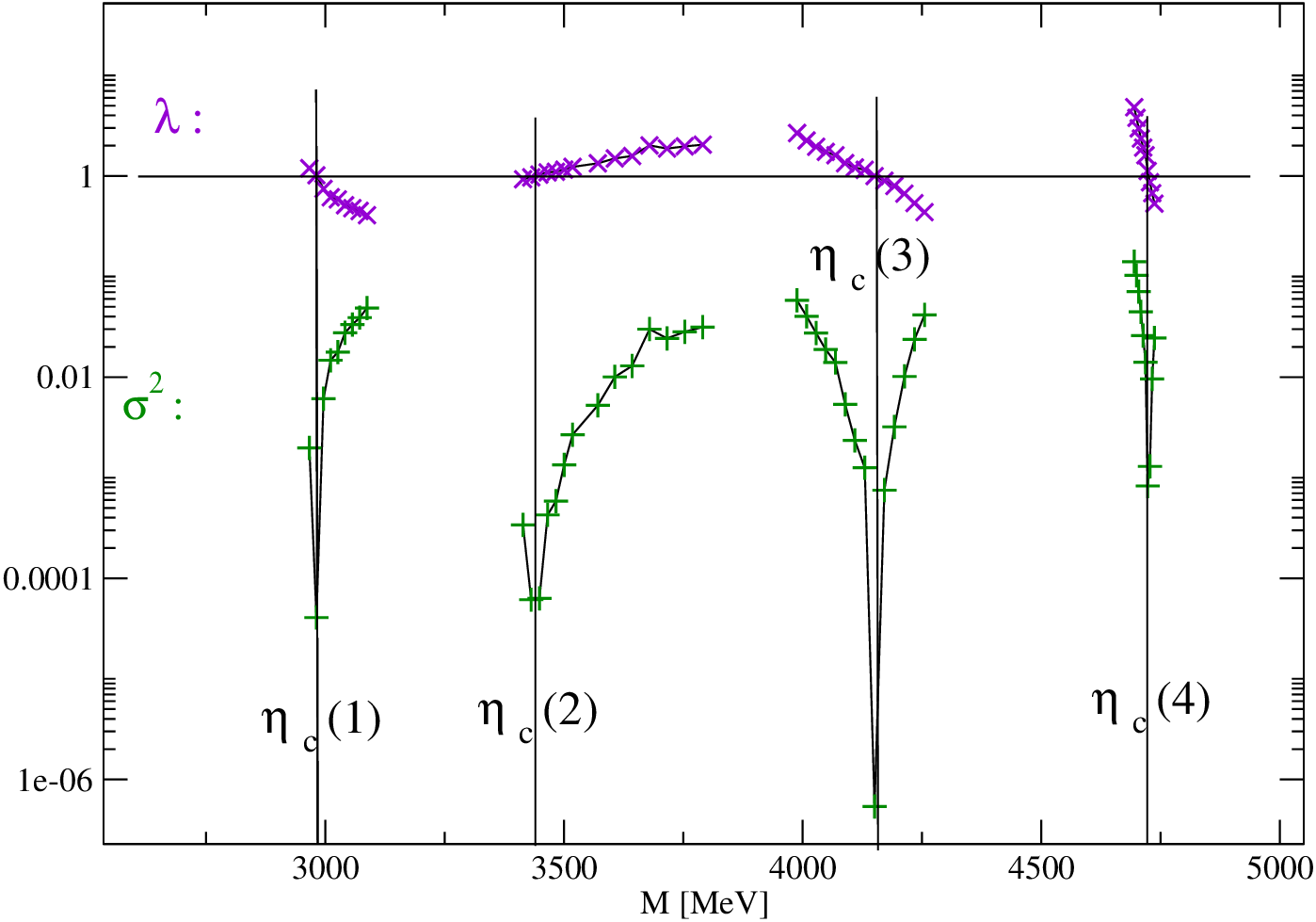}
\caption{The eigenvalue $\lambda$ and the numerical error $\sigma$ from the solution of BSE the  solution for the ground state and the the first three excited state of pseudoscalar  charmonium. The definitions can be find in \cite{saulipsy}, the bound states are for $\lambda \rightarrow 1$  $\sigma\rightarrow 0$ when satisfied simultaneously for the meson mass $M$.}
\label{tri}
\end{figure}

The obtained masses are listed in the Tab \ref{tabbse2} further predicted and  not yet observed excited $\eta_{c}$ charmonia states  we only list here: 5436,6186,7030MeV,...
Our calculations  ignore interaction with lighter quarks/meson
channels. Therefore  these states remain  narrow in our approximation and are  difficult to compare with  a broad resonance observed  experimentally.    We  show our ultimate  numerical search for  eigenvalue $\lambda$  and the iteration difference $\sigma^2$ (now for BSE)  in the  figure \ref{tri}. A single point shown in this figure required one day of heating a recent single-processor machine. 

\begin{table}[t]
\begin{center}
\small{
\begin{tabular}{|c|c|c|} \hline \hline 
BSE  & EXP. & HLC  \\
\hline \hline
2980 & 2980  & 3073\\
\hline
3442   & 3638 & 3679\\
\hline
4150 & --- & 4171\\
\hline
4720 & ---  & --- \\
\hline \hline
\end{tabular}}
\caption[99]{Calculated spectrum compared  with experimental data (second column). We also compare with the Light front hamiltonian approach results
   \cite{LMZV2015} in the third column).
\label{tabbse2}}
\end{center}
\end{table}

The resulting charm quark spectral functions are added into the fig. \ref{jedna} for comparison. 
Since the spectral functions are dimensinful objects, we introduce the dimensionless ones  $\sqrt(o)\sigma_s(o)$ and $o\sigma_v(o)$, for a better comparison of the spectral functions of different flavors. These objects are compared in the figure \ref{dva}. 
The main  rigid property is that the perturbative  on-shell singularity is washed out to into a broad peak,  and a heavy  a free quark  does not exist.
The confinement that manifests in the spectral framework of DSEs is the same for light and heavy quarks.
 The change in the kernel when transitioning from the light to the heavy meson  case reveals the importance of the quark-gluon vertex.

The spectra of bottomonia  has been obtained in a similar fashion \cite{sauliFBS}, but for the vectors $Y$ instead. However it  turns out to be more complicated due to the presence of infrared singularities, which are  unavoidable  for spin-one bound states.  
A more comprehensive study of the $\eta_B$ meson is planned for the future.

\section{Conclusion}

We studied the effect of the running quark mass on the properties of mesons. While its inclusion is necessary to describe the Goldstone character of the pion, we have shown that it is also remarkable for the heavy meson sector. To this end, we solved a coupled set of spectral quark Dyson-Schwinger and Bethe-Salpeter equations to obtain the masses of the desired mesons.
Notably, we obtained the masses of excited states of pseudoscalar charmonia. We avoided using relativistic versions or various stringy potentials believed to be "derivable" from gauge-invariant Wilsonian loops. These are unnecessary, as accounting for running quark masses in the equation for bound states fully substitutes their effect.  The form of  gauge invariant  quark propagator as has been established  in \cite{ACS2023}, is very interesting, but   far from  unique path  to observable hadroproductions. 

 The running coupling and the running of quark masses are unavoidable phenomena  stemming from quantum loop corrections in QCD.
Specifically, the latter phenomenon was examined in the calculation of   $\eta_c(N)$ radial excitations, where  the charm mass varied from $1$ to $1.5 GeV$. This  turns out to be quantitatively significant  and nonnegligible for charmonium physics.

As an added benefit, we obtained spectral quark functions when we solved the DSEs for the quark propagator. At this stage, we should note that there is no urgent need for a purely continuous spectral function, and the use of the constituent quark mass approximation (i.e., the delta function spectrum) provides almost indistinguishable results. Continuous spectral functions do not make the practical life of particle physicists easier; however, they offer a very simple picture of confinement for both light and heavy quarks.

Quarks never live on-shell, nor is their mass ever close to the on-shell mass. The sharp singularity of the perturbative/free quark propagator is completely washed out, and the physical threshold is not presented in a peculiar way. Instead, the imaginary part of the quark propagator gradually grows from the anomalous thresholds, or zero momenta. It goes without saying that the picture can become technically complicated if complex anomalous thresholds arise.

We use the ladder-rainbow approximation, and we acknowledge that the scheme was tuned to minimize deviations from spectrality.
Whether the simple picture obtained here survives further improved approximations or extension to other meson flavor sectors can be and will be checked in future studies.

 %%%%%%%%%%%%%%%%%%%%%%%%%%%%%%%%%%%%%%%%%%%%%%%

%
\end{document}